\documentstyle[aps,epsfig,multicol]{revtex}
\draft
\begin{document}
\title{Conductance oscillations in metallic nanocontacts }
\author{P. Havu, T. Torsti, M. J. Puska, and R. M. Nieminen}
\address{ Laboratory of Physics,
Helsinki University of Technology,
P.O. Box 1100, FIN-02015 HUT, Finland}

\maketitle
\begin{abstract}
We examine the conductance properties of a chain of Na atoms between
two metallic leads in the limit of low bias.  Resonant states
corresponding to the conductance channel and the local charge
neutrality condition cause conductance oscillations as a function of
the number of atoms in the chain. Moreover, the geometrical shape of
the contact leads influences the conductivity by giving rise
to additional oscillations as a function of the lead opening angle.
\end{abstract}


\begin{multicols}{2}
\section{Introduction}

The prospects to fabricate molecular devices, extending from a few
atoms to 10 $\mu$m long carbon nanotubes, have inspired an active
research field \cite{joachim}. The most modern experimental, theoretical and
computational methods are used to investigate and model ultimate-size
transistors, switches, gates, and memory devices as well as their
coupling with the rest of the electronic circuit. In this work we
concentrate to chains of Na atoms as ultimate conductors. Similar
atomic-scale metallic nanowires have been produced from Au by a
scanning tunneling microscope (STM) \cite{tippi,ohnishi} and by the
mechanically controllable break junction (MCBJ) technique
\cite{ruitenbeek}. The formation of Na nanowires  has been 
investigated by first-principles calculations \cite{barnett,hannu,muodostus}
and the ensuing conductances have been compared with experiments 
\cite{muodostus}.

The conductance of a nanowire shows nonmonotonic behavior as a
function of its length. However, the role of the contact leads cannot
be ignored, as shown by Sim {\em et al.} \cite{sim} and Yeyati {\em et al.}
\cite{levy}. The purpose of this article is to study the dependence of
the conductance not only on the length but also on the shape of the
contacts. We show that the shape of the contact, i.e. the bluntness of
the STM tip, influences the conductance in an oscillatory way.

Yeyati {\em et al.}  \cite{levy} recognized the importance of
resonance states to the conductance of sharp leads. These resonances
exist only at sharp tips, not at blunt ones. In the case of monovalent
atoms, such as Na or Au, the appearance of a resonance at the Fermi
level can explain the first quantized conductance step as a multiple
of $2e^2/h$. Sim {\em et al.}  \cite{sim} showed that the charge
neutrality required for the Na chain and the tips of the contacts
leads to an even-odd behavior in the conductance as a function of the
number of atoms in the chain. This is because for an odd number of
atoms the resonance is half-occupied whereas for an even number of
atoms the resonance is fully-occupied, leading to a low density of
states at the Fermi level.

As Sim {\em et al.} \cite{sim} we see the even-odd behavior for lead
tips. In addition, the conductance oscillates as a function of the lead
opening angle. This results from the change in the angular character
of electrons screening the tip regions of the leads.



\section{Model}

\begin{figure}[htb]
\begin{center}
\epsfig{file=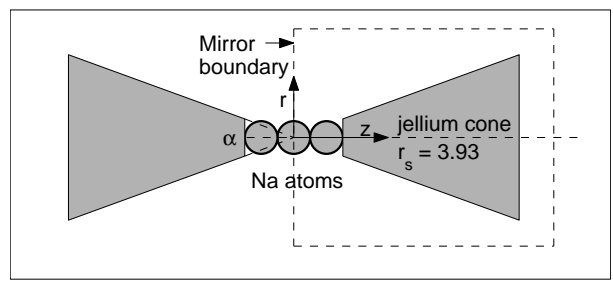,width=0.45\textwidth}
\end{center}
\caption{\label{kartio} Geometry of the present model. Na pseudoatoms 
are located between two jellium cones. The cone angle $\alpha$ can be
varied continuously. The dotted line gives the boundaries of the 
calculation cell. For the details, see also the text.}
\end{figure}

As in Ref \cite{sim}, we model the nanocontact by a chain of $N$ Na
pseudoatoms \cite{pseudo}. To enable a systematic study of the shape
effects, we model the leads using the ``stabilized jellium''
model \cite{jellium}, where leads of discrete ions have been smeared
out to continuous positive charge distributions and a constant
potential energy term has been added to stabilize the Na bulk electron
density. The computational geometry is depicted in
Fig. \ref{kartio}. The positive background charge of each cone
corresponds to $M$ electrons. The constant density $n_+$ of the
positive background is determined from the Na valence electron density
parameter $r_s$ = 3.93 a.u.  ($n_+ = 3/(4\pi r_s^3)$).  Near the atom
chain, the radius of the cones is $r_s$ and the larger radii are
determined by the number of electrons in the cones, which can be
varied at will. We solve the electronic states of the system using the
density-functional theory within the local density approximation
\cite{xc} assuming the limit of zero external bias.

The system has cylindrical symmetry, so that the problem is
computationally two-dimensional. In the middle of the atom chain there
is a mirror plane perpendicular to the symmetry axis. The electronic
wave functions are required to be even or odd with respect to the
mirror plane.  Naturally, the Poisson equation involved in the
self-consistent loop is solved by requiring mirror symmetry for the
Coulomb potential.  On the other boundaries, {\em i.e.}  on a plane
parallel to the mirror plane but far outside the jellium cone and on a
cylinder surface far from the axis, the wave functions are required to vanish.  The numerical solutions of the density-functional equations
are obtained by a real-space multigrid scheme \cite{mika}.

In the case of a chain of Na atoms there is only one conducting
eigenchannel contributing to the electron transport at the Fermi
level. Then the mirror symmetry and the Friedel sum rule give the
conductance $G$ in the zero-bias limit as \cite{datta}
\begin{equation}
G={2e^2 \over h} \sin ^2 \left [{\pi \over 2} (N_e - N_o) \right],
\label{friedel}
\end{equation}
where $N_e$ and $N_o$ are the numbers of electrons in the even and odd
parity states, respectively. In an infinite system, $N_e$ and $N_o$
are calculated by integrating the density of states up to the Fermi
level.  The states are assumed to have the spin degeneracy and to be
occupied accordingly. Then the difference of $N_e$ and $N_o$ arises
only from the electrons of the conducting eigenchannel, because only
their wave functions may have a finite value at the mirror
plane. In our model the conducting eigenchannel corresponds to the
quantum number $m$~=~0 of the angular momentum around the symmetry
axis.

The states with a finite value at the mirror plane ($m$~=~0) span an
energy region starting somewhat below the Fermi level. They are
strongly localized at the chain atoms and the lead tips. These
resonance states, as occupied up to the Fermi level, have to
participate in neutralizing the positive charge of the atom chain and
the tips of the leads. According to the calculations by Sim {\em et
al.} \cite{sim}, for Na atom chains between two Na (111) tips, the
neutrality requirement leads for an odd number $N$ of Na atoms to a
half-filled resonance in the conduction eigenchannel. This means that
the Fermi level is at the resonance state, which is filled in a
spin-compensated manner totally by one electron.  As a result, the
difference $N_e-N_o$ is odd and the conductance is close to
$2e^2/h$. For an even $N$ the Fermi level is off the resonance and
the conductance will be less than unity.  Below we will show that also
electron states which do not correspond to the conducting eigenchannel
($m>0$) participate in the charge neutralization in the tips,
modifying the conductance behaviour.

In order to study the convergence of the results as a function of the
system size we have performed calculations with $M$ = 200 and 300
electrons in each jellium cone. The finite size results in a discrete 
eigenvalue spectrum which is smoothed using the Fermi-Dirac
broadening with a finite temperature when occupying the electron
states. The broadening should be larger than the single electron level
spacing but smaller than the distance between the resonances. The
level spacing decreases as $\sim M^{-2/3}$ so that we can decrease the
temperature with increasing $M$ and obtain sharper resonance
structures. For 200 electrons we have chosen the temperature of
600 K, which was used also by Sim {\em et al.} \cite{sim} 
for cones of 60 and 95 electrons. In the case of 300 electrons in a
cone we choose the temperature of 400 K.

\section{Results}


Our main result is the finding that the conductance of a chain
of atoms depends on the cone angle of the leads. In order to
understand the reasons we first discuss the changes in the
electronic structure in terms of such integrated quantities as
the electron density and the local density of states (LDOS).
Thereafter the conductance, which reflects changes in the
electronic structure on a microscopic level, is given.   

\begin{figure}[htb]
\begin{center}
\epsfig{file=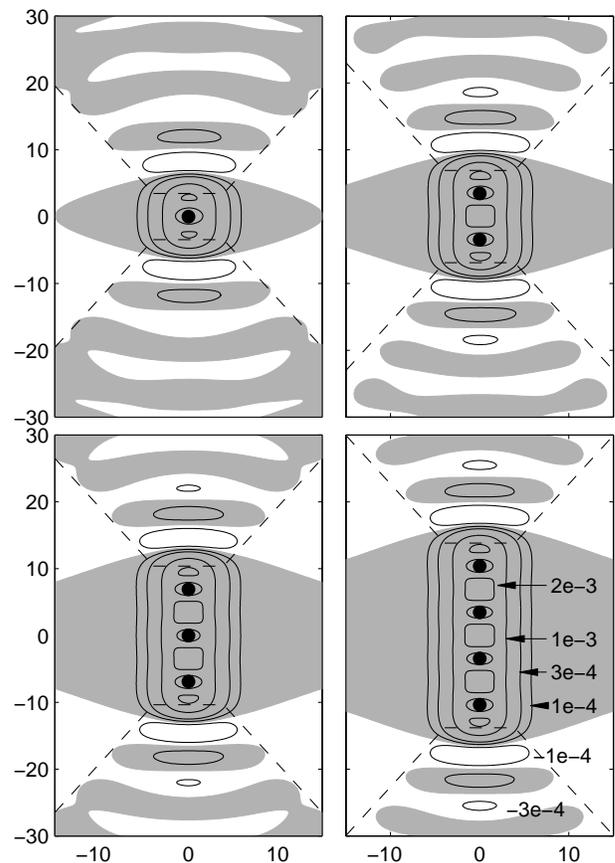,width=0.45\textwidth}
\end{center}
\caption{\label{density} Electron densities induced by chains of
one, two, three and four pseudoatoms between two jellium leads.
The cone angle is 69$^\circ$ and each of the cones contains 300 
electrons. The density values of the contours are given in
atomic units ($a_o^{-3}$). Areas of positive induced density are shown 
as gray and the ionic positions as black circles. }
\end{figure}

\begin{figure}[htb]
\begin{center}
\epsfig{file=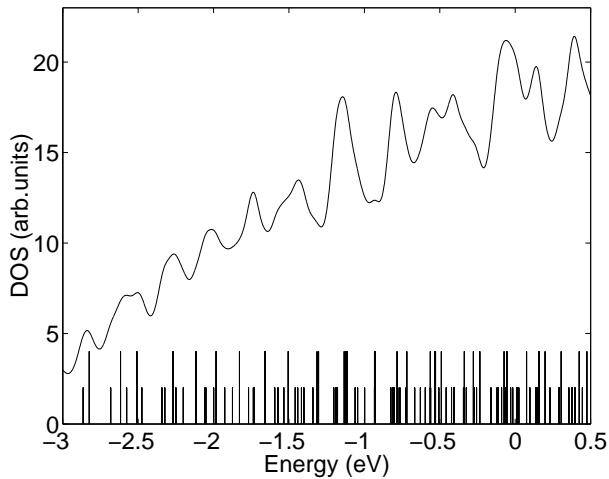,width=0.45\textwidth}
\end{center}
\caption{\label{ldosKoko} 
Chain of two Na atoms between two jellium leads with the cone angles of
69$^\circ$. The density of states (DOS) is given with respect to the 
Fermi level (energy zero). The single electron eigenenergies are marked 
with bars at the bottom of the figure. The $m=0$ states are denoted by
longer bars than the $m>0$ states.
}
\end{figure}

Fig. \ref{density} shows the electron density induced by chains of 
one to four Na atoms between two jellium leads containing 300 
electrons each. The cone angle is 69$^\circ$. The induced density 
is calculated as the difference between the whole system and the 
corresponding system consisting of the leads only. In accordance 
with the calculations \cite{sim} having 95 pseudoatoms as leads,
the main part of the induced density is localized at the atom 
chain and the density decays rapidly in the cones. This indicates
convergence as a function of the system size. For an even $N$ 
the electron density is slightly more localized than in the case 
of an odd $N$ because for an even $N$ strong resonance states 
are below the Fermi level and are fully occupied. The density of
states (DOS) of the system with two Na atoms is shown in Fig. 
\ref{ldosKoko}. It is obtained from the discrete energy level
spectrum given at the bottom of the figure by substituting
the lines by Lorentzians, the widths of which correspond to 
the temperature used in the self-consistent electronic structure 
calculations. We see that around the Fermi level the single 
electron level spacing is small in comparison with distances
between the strong DOS peaks. The latter are related 
to the $m=0$ resonance structure (the $m=0$ eigenenergies are
given in the figure as long vertical lines). It can be seen 
that the region of peaks with a larger amplitude reaches about 
1 eV below the Fermi level. 

\begin{figure}[htb]
\begin{center}
\epsfig{file=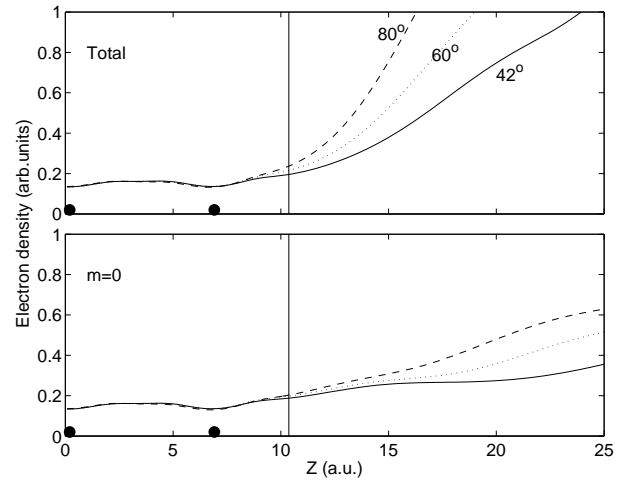,width=0.45\textwidth}
\end{center}
\caption{\label{AaltoInteg} Chain of two Na atoms between two jellium leads.
The electron density integrated over the plane perpendicular to the cylinder 
axis as a function of the position $z$ along the axis is given. The upper 
panel shows the total density whereas the lower panel gives the contribution
due to the $m=0$ states, only. Densities for the three cone angles of 
42$^\circ$, 60$^\circ$, and 80$^\circ$ are given by solid, dotted and 
dashed lines, respectively. The vertical line marks the 
jellium edge and the black circles denote the Na atom positions. }
\end{figure}

The character of the electron density near the tips of the leads
changes as the cone angle increases. This is demonstrated in
Fig. \ref{AaltoInteg} which shows the total electron density 
and its $m=0$ state component for three different cone angles.
The densities are integrated over a plane perpendicular to the 
cylinder axis and are given as a function of the position 
$z$ along the axis. Naturally, the total electron density 
increases close to the tip as the cone angle increases. But 
it is important to note that the $m>0$ 
electron states contribute most of the density increase. As 
a result, the $m=0$ resonance states feel an increased Coulomb
repulsion and are pushed upwards in energy. For the same 
reason, also other states with low $m$ ($>0$) values rise in
energy, but to a lesser extent.

\begin{figure}[htb]
\begin{center}
\epsfig{file=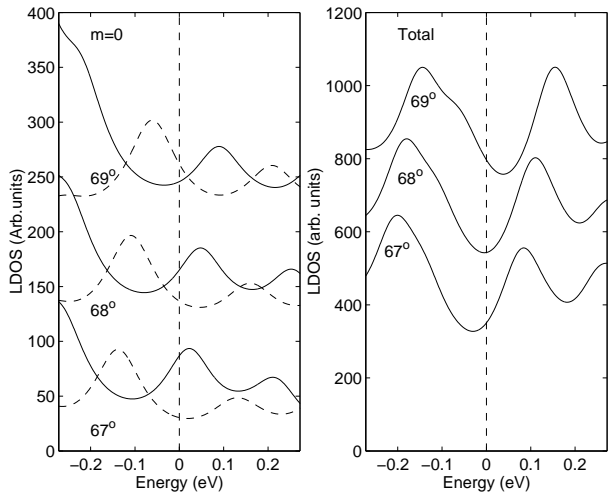,width=0.45\textwidth}
\end{center}
\caption{\label{LDOSMuutos} Chain of two Na atoms between two jellium leads.
The local density of states (LDOS) near the Fermi level (energy zero)
is given for different cone angles. The left panel shows the 
the even (solid line) and odd (dashed line) $m=0$ LDOS's calculated for 
the atom chain between the jellium edges. The right panel shows the
total LDOS's calculated for the atom chain and the tips of the
leads to the depth of $r_s$ from the jellium edge. The LDOS's
corresponding to the two uppermost angles are shifted in steps
of 100 and 200 units in the left and right panels, respectively.
}
\end{figure}

The shift of the $m=0$ states to higher energies can be seen in the
local density of states (LDOS) at the Na atom chain. We calculate the
LDOS similarly to the DOS discussed above, but now the area of a
Lorenzian peak is the density of the state in question integrated over
a volume limited by two planes perpendicular to the symmetry axis. In
the left panel of Fig. \ref{LDOSMuutos} we show the LDOS of the even
and odd $m=0$ states for the cone angles of 67$^\circ$, 68$^\circ$,
and 69$^\circ$. The integration is limited to the atom chain, {\em
i.e.}  between the planes at the central jellium edges. The LDOS of
the chain has a structure similar to the result of the atomistic lead
calculations \cite{sim}.  When we open the cone angle, the $m=0$
states are shifted, as discussed above, to higher energies. The
conductance of the chain is close to $2e^2/h$ when the Fermi level is
located on top of a peak of the $m=0$ LDOS. The left panel of
Fig. \ref{LDOSMuutos} indicates that the conductance varies rapidly as
a function of the cone angle. The right panel of Fig. \ref{LDOSMuutos}
gives the total LDOS calculated by including the tips of the cones
into the integration; the limiting planes are at the depth of $r_s$
from the central jellium edges. We see that in this volume, which is
important for the charge neutrality of the atom chain, the $m=0$
features of the left panel are hardly visible. The change of this
total LDOS is also less dramatic than that of the $m=0$ LDOS at the
atom chain.

\begin{figure}[htb]
\begin{center}
\epsfig{file=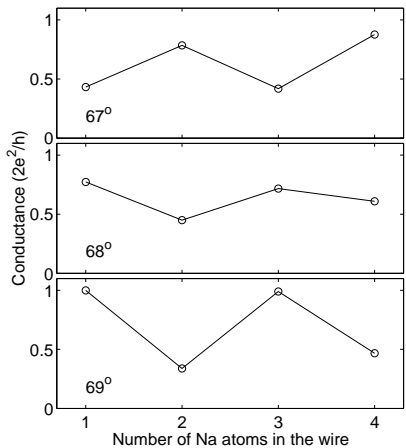,width=0.3\textwidth}
\end{center}
\caption{\label{Noscillations} Conductances of the chains  of
one, two, three and four Na atoms between two jellium leads.
The conductances are shown for three different jellium cone angles.
The number of electrons in each cone is 300.
}
\end{figure}

On the basis of Figs. \ref{AaltoInteg} and \ref{LDOSMuutos} we can
conclude that the $m>0$ states have an important contribution to the
local charge neutrality and influence the conductance of the
chain. This is seen in Fig.  \ref{Noscillations} which gives the
conductances of the chains of one to four Na atoms corresponding to
these three cone angles.  The even-odd oscillation as a function of
the number of chain atoms is clear as well as its gradual phase
change.

\begin{figure}[htb]
\begin{center}
\epsfig{file=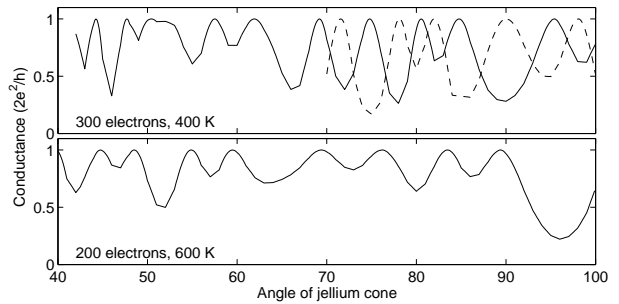,width=0.45\textwidth}
\end{center}
\caption{\label{Aoscillations} Chain of three Na atoms between two 
jellium leads. The conductance is given as function of cone angle
for systems of 200, and 300 electrons in one lead. 
The temperature used in occupying the one-electron states is
600 K for the smaller system and 400 K for the larger system.
The dotted line shows the conductance of the chain of two atoms. 
}
\end{figure}

Finally, Fig. \ref{Aoscillations} shows the conductance for the $N$ =
3 chain as a function of the cone angle. Results for system sizes
corresponding to 200 and 300 electrons in each jellium cone are given.
It can be seen that for both system sizes the conductance is a
regularly oscillating function of the cone angle. The temperature is
closely connected to the amplitude of the oscillations so that a lower
temperature gives stronger oscillations. If we choose a high enough
temperature, the oscillation as function of angle disappears but so
does the oscillation as function of $N$.  The number of oscillations
seems to increase as the system size increases. This is because the
resonance structure becomes more resolved including higher $m$
components as the size of the system increases.

The differences in the curves for the systems of 200 and 300 electrons
in Fig. \ref{Aoscillations} indicate that the finite system size has
still some influence. However, there are also clear similarities.  For
example, there appears shallow conductance minima between deeper
minima.  This structure is connected to the clustering of the
resonance $m=0$ energy levels as seen in the level spectrum at the
bottom of Fig. \ref{ldosKoko}. The effect can be considered as
reminiscent of the shell structure in the system of conical
confinement.

In Fig. \ref{Aoscillations} we show for the larger system the
conductance of the chain of two Na atoms at large cone angles. 
The conductances of the $N=2$ and $N=3$ chains are in opposite phases
and according to our calculations this state continues at least
to the angle of $110^\circ$.
Thus, we were not able to find a limiting blunt angle at which the
influence of the resonance states and the even-odd oscillations
in the conductance disappear \cite{sim,levy}. 

The oscillation of the conductance as a function of the cone angle
means the alternation of perfect electron transport and partial
electron scattering in the $m=0$ channel. The electron scattering is
caused by the interference between the $m=0$ channel and the
non-conducting $m>0$ channel. The situation is analogous to the
transport through a quantum wire with a side-coupled quantum dot
\cite{kang,torio}. If the Kondo resonance of the quantum dot lies at
the Fermi level the transport channel DOS at the Fermi level is
suppressed (the situation is also called as the Fano antiresonance)
and the conductance has a dip. In the present model, the $m>0$
channels play the role of the side-coupled quantum dot.

As discussed in the Introduction, atomic chains have been
experimentally observed for Au \cite{ohnishi,ruitenbeek}. These chains
have been interpreted to contain up to four Au atoms. It has been noted that
before breaking the last conductance plateau of about one conductance
quantum is not smooth but exhibits abrupt changes synchronized with
abrupt changes in the elongation force \cite{bahn}. The steps in the
conductance may be of the order of one tenth of the conductance
quantum. Our findings give two explanations for these small
conductance changes. The conductance may jump abruptly when the number
of atoms in the chain increases. On the other hand, the geometry of
the lead tips changes affecting also the conductance. These two
effects may be superimposed during the actual elongation process.



\section{Conclusions}

We have studied the electronic properties of chains of Na atoms
between two metallic leads using a pseudoatom-jellium model. The
conductances of the chains are calculated using the Friedel sum rule.
The conductances show even-odd oscillation as function of number of
atoms in the chain confirming the earlier results obtained by a purely
atomistic model \cite{sim}. There is another oscillation: the
conductance oscillates as a function of the lead cone angle. The
even-odd oscillation is a direct result of the resonance structure of
the electron states of the conducting channel and the charge
neutrality at the atom chain and the tips of the leads. The
conductance oscillations as a function of the cone angle reflect the
fact that also electrons of the non-conducting channel take part in
the charge neutralization and that the balance between the conducting
and non-conducting channel electrons depends on the cone angle. Thus,
when determining the conductance of a nanoconstriction it is important
to solve self-consistently for the electronic structure, including the
lead regions.

\acknowledgments

We acknowledge the generous computer resources from the Center for
Scientific Computing, Espoo, Finland.  This research has been
supported by the Academy of Finland through its Centers of Excellence
Program (2000-2005).


\end{multicols}
\end{document}